\documentclass[sigconf]{acmart}
\usepackage{booktabs} 
\usepackage{enumitem}
\usepackage{makecell}
\usepackage{textcomp}
\usepackage{amsmath}
\usepackage{algorithm}
\usepackage[noend]{algpseudocode}
\usepackage{pdfpages}
\usepackage{graphicx}
\usepackage{amssymb}
\usepackage{mathrsfs}
\usepackage{xcolor}
\usepackage{epsfig}
\usepackage{graphicx}
\usepackage{multirow}

\usepackage{enumerate}
\usepackage{balance}

\usepackage{lineno}

\copyrightyear{2020}
\acmYear{2020}
\setcopyright{acmlicensed}
\acmConference[KDD '20]{Proceedings of the 26th ACM SIGKDD Conference on Knowledge Discovery and Data Mining}{August 23--27, 2020}{Virtual Event, CA, USA}
\acmBooktitle{Proceedings of the 26th ACM SIGKDD Conference on Knowledge Discovery and Data Mining (KDD '20), August 23--27, 2020, Virtual Event, CA, USA}
\acmPrice{15.00}
\acmDOI{10.1145/3394486.3403258}
\acmISBN{978-1-4503-7998-4/20/08}

\begin{document}
\fancyhead{}

\title{Interactive Path Reasoning on Graph for Conversational Recommendation}

\author{Wenqiang Lei$^1$, Gangyi Zhang$^2$, Xiangnan He$^2$$^*$, Yisong Miao$^1$, Xiang Wang$^1$, Liang Chen$^3$, Tat-Seng Chua$^1$}

\thanks{$^*$Xiangnan He is the Corresponding Author}

\affiliation{\institution{$^1$National University of Singapore, $^2$University of Science and Technology of China,
$^3$Sun Yat-Sen University}}

\email{wenqianglei@gmail.com, gangyi.zhang@outlook.com, xiangnanhe@gmail.com, miaoyisong@gmail.com}
\email{xiangwang@u.nus.edu, chenliang6@mail.sysu.edu.cn, chuats@comp.nus.edu.sg}



\begin{abstract}

Traditional recommendation systems estimate user preference on items from past interaction history, thus suffering from the limitations of obtaining fine-grained and dynamic user preference. 
Conversational recommendation system (CRS) brings revolutions to those limitations by enabling the system to directly ask users about their preferred attributes on items. However, existing CRS methods do not make full use of such advantage --- they only use the attribute feedback in rather implicit ways such as updating the latent user representation.
In this paper, we propose \textbf{C}onversational \textbf{P}ath \textbf{R}easoning (\textbf{CPR}), a generic framework that models conversational recommendation as an interactive path reasoning problem on a graph. It walks through the attribute vertices by following user feedback, utilizing the user preferred attributes in an explicit way. By leveraging on the graph structure, CPR is able to prune off many irrelevant candidate attributes, leading to better chance of hitting user preferred attributes.
To demonstrate how CPR works, we propose a simple yet effective instantiation named \textbf{SCPR} (\textbf{S}imple \textbf{CPR}). 
We perform empirical studies on the multi-round conversational recommendation scenario, the most realistic CRS setting so far that considers multiple rounds of asking attributes and recommending items.
Through extensive experiments on two datasets Yelp and LastFM, we validate the effectiveness of our SCPR, which significantly outperforms the state-of-the-art CRS methods EAR~\cite{lei20estimation} and CRM~\cite{Sun:2018:CRS:3209978.3210002}. 
In particular, we find that the more attributes there are, the more advantages our method can achieve.

\end{abstract}

\maketitle

\section{Introduction}
\label{sec:introduction}

Personalized recommendation systems have been standard fixtures in many scenarios like E-commerce (e.g., Amazon) and content sharing platforms (e.g., YouTube). They traditionally conduct recommendations by inferring user preference on items from their historical actions~\cite{NCF,FM}. While proven to be a success, traditional methods suffer from the intrinsic limitation of passively acquiring user feedback in the process of making recommendations. Such information asymmetry makes it hard to obtain dynamic and fine-grained user preference, preventing the system to provide accurate and explainable recommendation service. 

The recently emerging conversational recommendation system (CRS) brings revolutions to the aforementioned limitation.
CRS is envisioned as the deep composition of a conversational system and a recommendation system~\cite{lei20estimation}. It makes recommendations when interacting with users using natural languages and can proactively ask a user whether he/she likes an item attribute
or not. As such, CRS has the natural advantage of conducting dynamic and explainable recommendation by utilizing the user's preferred attributes as interpretable reasons. However, existing works only utilize attribute feedback implicitly by mapping attributes into a latent space, which we believe does not make full use of the advantage of attribute feedback. 
For example,
\citet{bi2019conversational, zhang2019toward} update the opaque user embedding once obtaining the user feedback on an attribute. \citet{lei20estimation} feed the preferred attribute into a variant of factorization machine~\cite{FM} to score items in the latent space. \citet{Sun:2018:CRS:3209978.3210002} feed the user attribute preference to a policy network, which is trained to decide the next action --- whether to make recommendations or ask an attribute.

\begin{figure}[t]
    \centering
	\includegraphics[width=8cm]{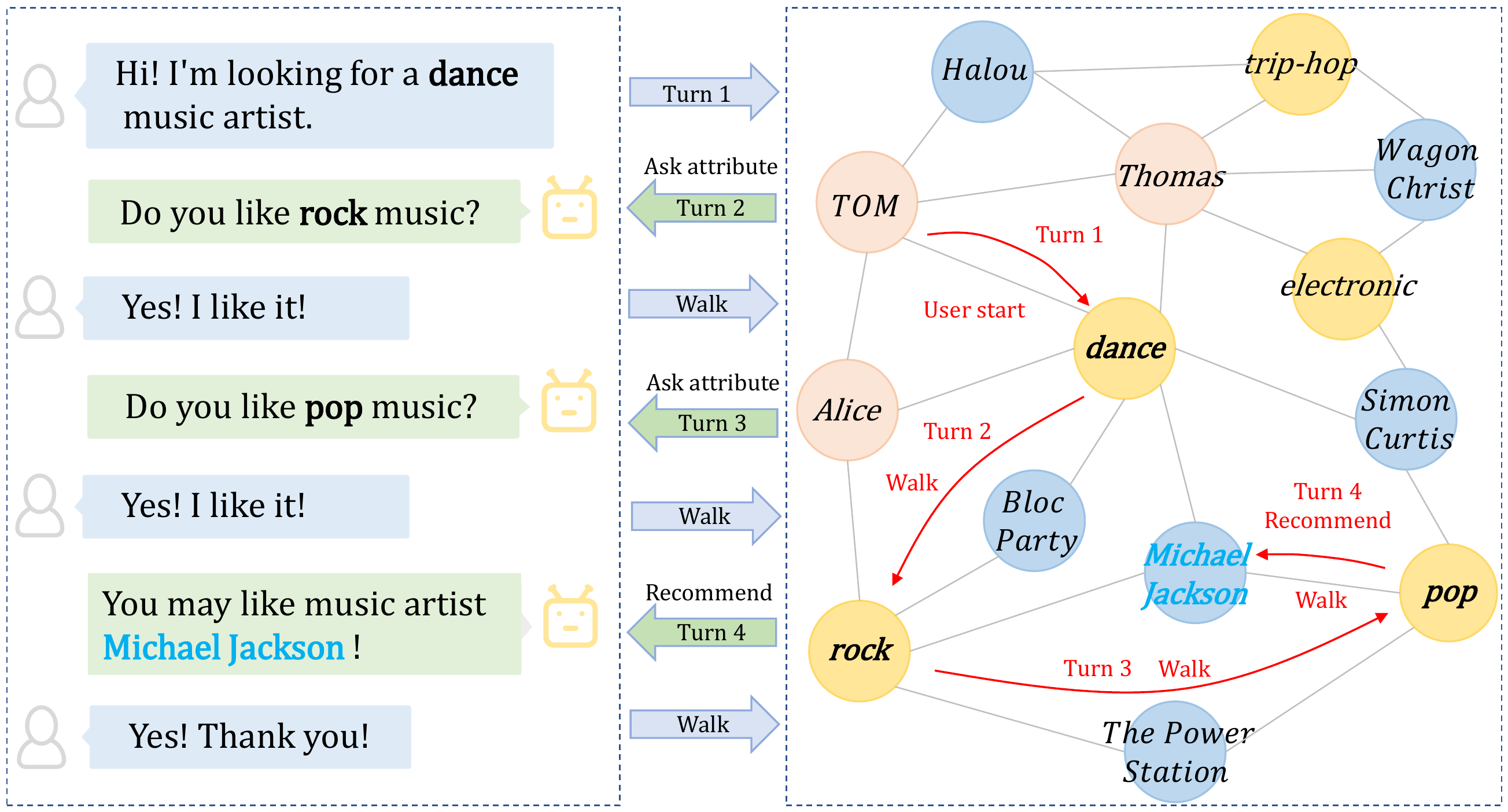}
    \caption{An illustration of interactive path reasoning in CPR. As the convention of this paper, light orange, light blue, and light gold vertices represents the user, attribute and items respectively. For example, the artiest \emph{Michael Jackson} is an item and and the attributes are \emph{rock}, \emph{dance} etc.}
    \label{fig:intro-exa}
    \vspace{-13pt}
\end{figure}

The key hypothesis of this work is that, a more explicit way of utilizing the attribute preference 
can better carry forward the advantages of CRS --- being more accurate and explainable. 
To this end, we propose a novel conversational recommendation framework called \textbf{C}onversational \textbf{P}ath \textbf{R}easoning (CPR). Inspired by the recent success of graph-based recommendation~\cite{wang2019neural}, we model conversational recommendation as the process of finding a path in user-item-attribute graph interactively. Figure~\ref{fig:intro-exa} shows an illustrative example. 
The vertices in the right graph represent users, items and attributes as well as other relevant entities. An edge between two vertices represent their relation, for example, a user-item edge  indicates that the user has interacted with the item, and a user-attribute edge indicates that the user has affirmed an attribute in a conversation session. 
A conversation session in our CPR is expressed as a walking in the graph. It starts from the user vertex, and travels in the graph with the goal to reach one or multiple item vertices the user likes as the destination. Note that the walking is navigated by users through conversation. This means, at each step, a system needs to interact with the user to find out which vertex to go and takes actions according to user's response.

We now go through an example in Figure \ref{fig:intro-exa} to better understand the process. A user $TOM$ is seeking a recommendation of music artists. The walking starts from the user vertex~(``$TOM$''), and the session is initialized by the user-specified attribute~(``$dance$''). Accordingly, the system makes its first step from ``$TOM$'' to ``$dance$''. Afterwards, the system 
identifies an \emph{adjacent attribute} (\emph{c.f.} Sec~\ref{sec:CPR_framework}) vertex on the graph to consult the user, or recommendation a list of items.
If the user confirms his preference to the asked attribute, the system will transit to that attribute vertex.
However, if the user rejects the attribute, or rejects a recommendation, 
the system will stay at the same vertex and consult the user for another attribute.
The session will repeat such cycle multiple times until the recommended items are accepted by the user\footnote{In our descriptions on graphs, we sometime directly use the word \emph{item}, \emph{attribute} or \emph{user} to refer to their corresponding vertices for simplicity.}. 

The proposed CPR framework, as a new angle of conducting conversational recommendation, conceptually brings several merits to the development of CRS:
\begin{itemize}[leftmargin=1mm]
    \item[1.] It is crystally explainable. It models conversational recommendation as an interactive path reasoning problem on the graph, with each step confirmed by the user. Thus, the resultant path is the correct reason for the recommendation. This makes better use of the fine-grained attribute preference than existing methods that only model attribute preference in latent space such as~\cite{lei20estimation}.
    \item[2.] It facilitates the exploitation of the abundant information by introducing the graph structure. By limiting the candidate attributes to ask as adjacent attributes of the current vertex, the candidate space is largely reduced, leading to a significant advantage compared with existing CRS methods like ~\cite{lei20estimation,Sun:2018:CRS:3209978.3210002} that treat almost all attributes as the candidates.
    \item[3.] It is an aesthetically appealing framework which demonstrates the natural combination and mutual promotion of conversation system and recommendation system. On one hand, the path walking over the graph provides a natural dialogue state tracking for conversation system, and it is believed to be efficient to make the conversation more logically coherent~\cite{jin2018explicit,acl18/sequicity}; on the other hand, being able to directly solicit attribute feedback from the user, the conversation provides a shortcut to prune off searching branches in the graph. 
\end{itemize}

To validate the effectiveness of CPR, we provide a simple yet effective implementation called \textbf{SCPR} (\textbf{S}imple \textbf{CPR}), targeting at the multi-round
conversational recommendation (MCR) scenario (\emph{c.f.} Sec~\ref{sec:pre}). 
We conduct experiments on the Yelp and LastFM datasets, comparing SCPR with state-of-the-art CRS methods \cite{lei20estimation,Sun:2018:CRS:3209978.3210002} which also use the information of user, item and attribute but does not use graph.
We analyze the properties of each method under different settings, including different types of questions (binary and enumerated) and different granularity of attributes.
We find that SCPR outperforms existing methods on recommendation success rate, especially in the settings where the attribute space is larger.

In summary, our contributions are two-folds:
\vspace{-5pt}
\begin{itemize}[leftmargin=1mm]
    \item We propose the CPR framework to model conversational recommendation as a path reasoning problem on a heterogeneous graph which provides a new angle of building CRS. 
    To the best of our knowledge, it is the first time to introduce graph-based reasoning to multi-round conversational recommendation.
    \item To demonstrate the effectiveness of CPR, we provide a simple instantiation SCPR, which outperforms existing methods in various settings. We find that, the larger attribute space is, the more improvements our model can achieve.
\end{itemize}

\section{Related Work}\label{sec:related}

The success of a recommendation system hinges on offering the relevant items of user interest accurately and timely. At beginning, recommendation systems are largely built on the collaborative filtering hypothesis to infer a distributed representation of the user profile. Representative models include matrix factorization~\cite{fastMF} and factorization machines~\cite{FM,NFM}. However, by nature, these approaches suffer from two intrinsic problems. The first one is the inability of capturing user dynamic preferences with the strict assumption that a user's interest is static over the long-term horizon~\cite{song2019session}. The second problem is the weak explainability as the user preference representation is only a continuous vector. Later works try to introduce Markov models~\cite{rendle2010factorizing} and multi-arm bandit methods~\cite{wu2018learning} to solve the dynamic problem but the explainability still remains to be unsatisfactory.
\par

Recently, \textbf{Graph-based recommendation methods} attract increasing research attention. One line of research leverages on the better expressiveness of the graph. They either explore implicit properties like collaborative signals~\cite{zheng2018spectral,wang2019neural} from the global connectivities, or focus on yielding better representations of user/items by incorporating latent network embeddings~\cite{yang2018hop}. Another line of work leverages on the explainability of the graph, modeling recommendation as a path reasoning problem on the graph. They aim to find a path from a user vertex to the target item, and use the resultant path as the recommendation reason~\cite{wang2019explainable,xian2019reinforcement}. While being explainable, such methods suffer from two problems: 1) they are still static models which intrinsically cannot capture the preference dynamics, and 2) the modeling complexity is high such that pruning becomes a critical step~\cite{xian2019reinforcement}.

\textbf{Conversational recommendation system} (CRS) becomes an appealing solution to both the dynamic preference and weak explainability problems as it dynamically gets user explicit feedback. As an emerging topic, various problems under different settings have been explored~\cite{zhang2019text, christakopoulou2016towards,nips18/DeepConv,Liao:2018, christakopoulou2018q,zhang2018towards,Sun:2018:CRS:3209978.3210002, priyogi2019preference, yu2019visual, sardella2019approach, zhang2019toward,chen-etal-2019-towards,  li2020seamlessly}, such as natural language understanding and generation~\cite{nips18/DeepConv,chen-etal-2019-towards}, multi-model and multi-media~\cite{Liao:2018}, monitoring user feedback on viewing, clicking and commenting~\cite{yu2019visual}, and attribute prediction~\cite{priyogi2019preference}.

We believe that how to dynamically ask attribute questions and make recommendations upon attribute answer is the key at current stage of conversational recommendation. 
As such, we consider the system asking user preference on attributes and making recommendation based on those attributes in a multi-turn basis. As discussed in Section~\ref{sec:introduction}, main works~\cite{bi2019conversational, zhang2019toward,lei20estimation, Sun:2018:CRS:3209978.3210002} along this line do not use attributes explicitly. We argue more explicitly utilizing the attribute would better carry forward the advantage of conversational recommendation. Therefore, this paper makes a key contribution to introduce graph to increase the explainability.

\begin{table}[t]
\caption{Main notations used in the paper.}\small
\vspace{-8pt}
\label{tab:notation}
\begin{tabular}{|m{1.5cm}<\centering|m{6cm}|}
\hline
$u,v,p$  & User, item, and attribute  \\ \hline
$P$  & An \textit{active} attribute path in the graph
\\ \hline
$aa_t$ & An adjacent attribute of the attribute $p_t$  \\ \hline
$\mathcal{AA}_t$ & The set of adjacent attributes of the attribute $p_t$  \\ \hline
$\mathcal{P}_{u}$ & The set of attributes confirmed by $u$ in a session \\ \hline
$\mathcal{P}_{cand}$ & The set of candidate attributes \\ \hline
$\mathcal{V}_{p}$ & The set of items that contain the attribute $p$ \\ \hline
$\mathcal{V}_{cand}$ & The set of candidate items;
\\ \hline
$a$ & The action of CPR, either $a_{ask}$ or $a_{rec}$
\\ \hline
\end{tabular}\vspace{-10pt}
\end{table}

\section{Multi-round Conversational Recommendation Scenario}\label{sec:pre}

As conversational recommendation is an emerging research topic, various settings have been explored in recently. 
This paper follows the \textit{multi-round conversational recommendation} (MCR) scenario since it is the most realistic setting in research so far~\cite{lei20estimation}. In a MCR setting, a CRS is free to ask attributes or make recommendation multiple times. We use a \emph{round} to emphasize one trial of recommendation. This is in contrast to the \emph{single-round conversational recommendation} as adopted by~\cite{Sun:2018:CRS:3209978.3210002} where the system asks attribute multiple times followed by making recommendation only once, after which the conversation session ends regardless of whether the recommendation succeeds. The \emph{multi-round} setting is more challenging than the \emph{single-round} one as a CRS has more freedom to take actions which makes the policy space more complex.

Specifically, an item $v$ is associated with a set of attributes $\mathcal{P}_{v}$. The attributes broadly cover various descriptions as long as it can describe certain properties of an item. For example, in the music artist recommendation domain (e.g., in the lastFM dataset), an item is a music artist and the attribute may be descriptions like \emph{Jazz}, \emph{Classic}, \emph{Energetic}, \emph{Peaceful} etc. The items and attributes are provided by the dataset. During a conversation session, a CRS obtains the user's fine-grained preference by asking whether he likes particular attributes. Based on such conversations, a CRS aims to provide accurate recommendations in the shortest conversational turns.

A conversation session starts on the user side, which initializes the attribute $p_{0}$ by specifying an attribute the user likes (e.g., I like some \emph{dance} music). 
Next, the CRS is free to ask his preference on an attribute selected from the candidate attribute set $\mathcal{P}_{cand}$ or recommend items from the candidate item set $\mathcal{V}_{cand}$. Then, the user needs to give feedback accordingly, either accepting or rejecting them. The CRS makes use of such feedback from the user --- if the user accepts the asked attribute, the CRS puts it in the preferred attribute set $\mathcal{P}_{u}$ and removes it from $\mathcal{P}_{cand}$. Then the CRS updates $\mathcal{V}_{cand}$ to $\mathcal{V}_{cand} \cap \mathcal{V}_{p}$, representing
the items containing all attribute confirmed by the user in the session. 
$\mathcal{V}_{p}$ denote the items containing the attribute $p$; if he rejects the asked attribute, the CRS removes it from $\mathcal{P}_{cand}$; 
if he rejects the recommended items, the CRS removes them from $\mathcal{V}_{cand}$.  Based on the updated sets, the CRS takes the next action, i.e., \emph{asking} or \emph{recommending}, and repeats the above process. The conversation session ends until the CRS hits the user preferred items or reaches the maximum number of turns $T$. 
This process is detailed in Algorithm~\ref{algo:MCR}. 

\renewcommand{\algorithmicrequire}{\textbf{Input:}}
\renewcommand{\algorithmicensure}{\textbf{Output:}}
\algnewcommand{\algorithmicand}{\textbf{ and }}
\algnewcommand{\algorithmicor}{\textbf{ or }}
\algnewcommand{\OR}{\algorithmicor}
\algnewcommand{\AND}{\algorithmicand}
\setlength{\textfloatsep}{1.5pt}
\begin{algorithm}[t]  
	\caption{\textbf{The MCR Scenario}}			
	\begin{algorithmic}[1]	
	\Require		
		user $u$,
        all attributes $\mathcal{P}$,
        all items $\mathcal{V}$,
        the number of items to recommend $k$,
        the maximum number of turns $T$;
	\Ensure recommendation result: success or fail;
    \State User $u$ specifies an attribute $p_{0}$; 
    \State Update:  $\mathcal{P}_{u}=\{p_{0}\}$;
    $\mathcal{P}_{cand}=\mathcal{P} \setminus p_{0}$;
    $\mathcal{V}_{cand}=\mathcal{V}_{p_{0}}$
    \For{turn $t=1,2,3...T$}
        \State Select an action $a$
        \If{$a == a_{ask}$}
            \State Select the top attribute $p$ from $\mathcal{P}_{cand}$
            \If{$u$ accepts $p_{t}$}
                \State Update: $\mathcal{P}_{u}=\mathcal{P}_{u} \cup p$;  $\mathcal{V}_{cand} =  \mathcal{V}_{cand} \cap \mathcal{V}_{p}$
            \EndIf
                \State Update: $\mathcal{P}_{cand} = \mathcal{P}_{cand} \setminus p$
            
        \Else[{$a == a_{rec}$}]
            \State Select the top-$k$ items $\mathcal{V}_{k}$ from $\mathcal{V}_{cand}$
            \If{User accepts $\mathcal{V}_{k}$}
                \State \emph{ Recommendation succeeds}; Exit.
            \Else[User rejects $\mathcal{V}_{k}$]
                \State Update: $\mathcal{V}_{cand} = \mathcal{V}_{cand} \setminus \mathcal{V}_{k}$
            \EndIf
        \EndIf
    \EndFor
    \State \emph{Recommendation fails}; Exit. 
	\end{algorithmic}
	\label{algo:MCR}
\end{algorithm}

Following \citet{lei20estimation}, it is noticeable that the above MCR scenario makes two assumptions. (1) It assumes that the user clearly expresses his preferences by specifying attributes without any reservations, and the items containing the preferred attributes are enough in the dataset. 
Given this assumption, the CRS takes the attributes accepted by the user as a strong indicator. For example, it only considers all items containing all attributes he accepts (line {\color{black}2} and line {\color{black}8} in Algorithm~\ref{algo:MCR}). This is because the items that contain all the preferred attributes have higher priority than the items do not. 
Since such higher-prioritized items are enough, ignoring the other candidate items is a reasonable simplification to this problem. 
(2) It assumes that the CRS does not handle strong negative feedback. This means, if a user rejects the asked attribute, the CRS does not distinguish whether the user \emph{does not care} it or \emph{hates} it. It is because such negative feedback is hard to obtain in current data, making it difficult to simulate in experimental surroundings. Therefore, the CRS equally treats all rejected attributes as \emph{does not care} and only removes the attributes from the candidate set without further actions like removing all items that contain the rejected attributes.

In this scenario, \citet{lei20estimation} distills several key research problems, such as:
(1) Which items to recommend?
(2) Which attribute to ask?
(3) When to ask attributes and when to make recommendations?
We next articulate how our method conceptually brings benefits to address these questions.

\begin{figure}[t]
    \label{fig:model-figure}
    \centering
	\includegraphics[width=9cm]{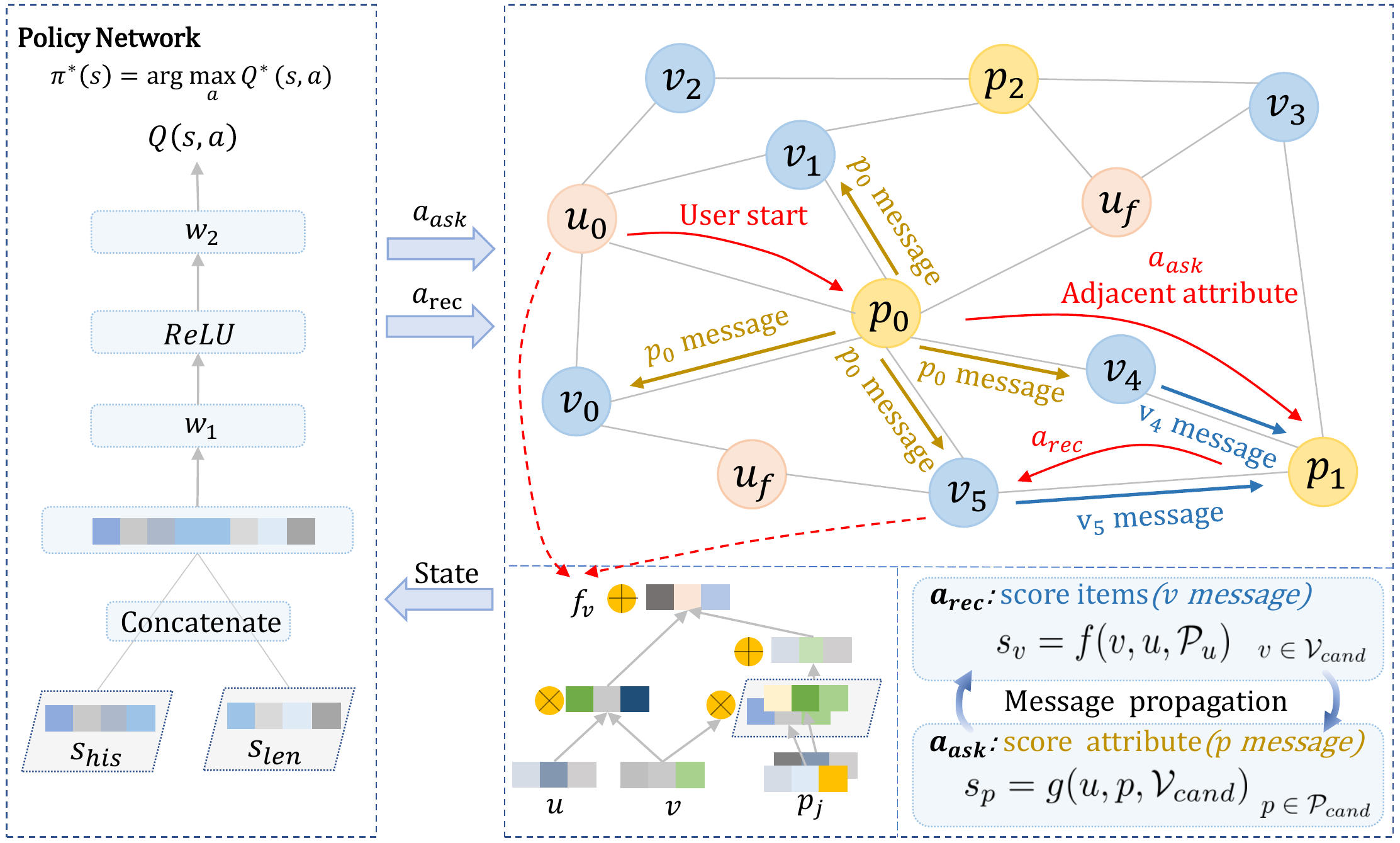}
    \caption{CPR framework overview. It starts from the user $u_0$ and walks over adjacent attributes, forming a path (the red arrows) and eventually leading to the desired item.
    The policy network (left side) determines whether to ask an attribute or recommend items in a turn. Two reasoning functions $f$ and $g$ score attributes and items, respectively.
    }
    \label{fig:model-figure}
\end{figure}

\section{Proposed Methods}
\label{sec:method}

We first propose \textit{Conversational Path Reasoning} (CPR), a general solution framework for graph-based conversational recommendation.  We then introduce a simple yet effective instantiation SCPR to demonstrate how it works. 

\subsection{CPR Framework}
\label{sec:CPR_framework}
A graph uses vertices to represent entities and edges to represent the relationships between entities. 
Specifically, a graph $G$ is defined as a set of triplets $\{(h,r,t)\}$, indicating a certain relation $r$ exists between the head entity $h$ and the tail entity $t$. 
In this paper, we consider the graph containing three types of entities, namely, user $u$, item $v$, and attribute $p$. 
The relations between each types of entities can vary a lot depending on specific datasets ( {\it c.f.} Table \ref{tab:basic} in Appendix~\ref{append-1}). For example, in Figure~\ref{fig:model-figure}, the edge between the $u_0$ and $p_0$ means the $u_0$ has specified his preference on attribute $p_0$ in his static profile; the edge between $u_0$ and $v_0$ indicate the user $u_0$ has interacted with $v_0$. Note that, in this paper, we do not specifically model different semantics of relations, and only care care whether there is an edge between two vertices for simplicity. In addition, the item, attributes and user information and their relations are also used by existing conversational recommendations systems~\cite{lei20estimation,Sun:2018:CRS:3209978.3210002}. The difference is that, our CPR organizes such three types of information in graph and leverages on the advantages of graph structure to conduct conversational recommendation.

In the MCR scenario, the system treats attributes as the preference feedback. To explicitly utilize these feedback, CPR performs the walking (i.e., reasoning) over the attribute vertices. 
Specifically, CPR maintains an active path $P$, comprising the attributes confirmed by a user (i.e., all attributes in $\mathcal{P}_{u}$) in the chronological order, and exploring on the graph for the next attribute vertex to walk. Note that, (1) CPR does not visit the attributes that have been visited before and does not consider the directions of edges.(2) The walking in CPR differs from existing work of graph-based recommendation~\cite{xian2019reinforcement,wang2019explainable}, which performs the walking over all types of vertices. 
We believe that restricting walking on attributes as in CPR brings two benefits. First, it emphasizes the importance of the attributes as explicit reasons for recommendation. Second,  it makes the walking process more concise, eliminating the uncertainty in an unnecessarily long reasoning path which might lead to error accumulation~\cite{xian2019reinforcement}.

Now, we move to the detailed walking process in CPR.
Assume the current active path is $P={p_0, p_1, p_2, ..., p_t}$. The system stays at $p_t$ and is going to find the next attribute vertex to walk.  
This process can be decomposed into three steps: \emph{reasoning}, \emph{consultation} and \emph{transition}.

\subsubsection{Reasoning} 
This is the beginning of a turn. It is triggered when an attribute is initialized or confirmed by the user. In this step, CPR scores items and attributes, solving the problem of \emph{which items to recommend} and \emph{which attribute to ask}.
In the context of MCR, CPR makes the key contribution of formalizing the scoring as message propagation on the graph. Because the scoring of attributes and items are interdependent, we adopt an alternating optimization strategy 
to optimize them in an asynchronous manner which has been proven to be effective~\cite{zhou2011functional}. 

First, the alternating optimization propagates messages from attributes to items to score the items (the light gold arrows in Figure~\ref{fig:model-figure}). Specifically, all attributes in the path $P$ (i.e., $\forall p_i \in \mathcal{P}_{u}$) together with the user vertex $u$ propagate messages to candidate items in $\mathcal{V}_{cand}$ (from Section~\ref{sec:pre}, we know that $\mathcal{V}_{cand}$ actually corresponds to the vertices directly connecting all $\mathcal{P}_{u}$). As an example, in Figure~\ref{fig:model-figure}, when a user initializes his preferred attribute $p_0$ (i.e., $P=p_0$), the CPR propagates messages from $p_0$ to its directly connected items (i.e., $v_0$, $v_1$, $v_4$, $v_5$) to score these items. The scoring function for each item can be any implementation of traditional recommender models, abstracted as
\begin{equation}\label{eq:s_v}
    s_v=f(v,u,\mathcal{P}_{u}),
\end{equation}
where $s_v$ is a scalar indicating the recommendation score of item $v$ in the current conversation session, and $\mathcal{P}_{u}$ denotes the attributes confirmed by $u$ in the session. 

Second, the candidate items in turn propagate messages to the candidate attributes (the light blue arrows in Figure~\ref{fig:model-figure}). The idea is that, with updated scores (i.e., $s_v$) calculated in the first step, the items provide additional information to find proper attributes to consult the user. For example, the attributes that can reduce the uncertainty in item scoring. 
Specifically, CPR leverages on the natural constraint of graph structure, considering only the transition to the \textbf{adjacent attributes} --- if the shortest path between attribute $p_t$ and $aa_t$ does not contain any other attribute, then $aa_t$ is the \textit{adjacent attribute} of $p_t$. 
For example, in the graph of Figure~\ref{fig:model-figure}, both $p_1$ and $p_2$ are the adjacent attribute of $p_0$. 
Formally, in CPR, the candidate attribute set $\mathcal{P}_{cand}=\mathcal{AA}_t \setminus (\mathcal{P}_{u} \cup \mathcal{P}_{rej})$, where $\mathcal{AA}_t$ stores all adjacent attributes of $p_t$ and $\mathcal{P}_{rej}$ is the attributes rejected by the user.
Finally, for a candidate attribute $p \in \mathcal{P}_{cand}$, 
its score is calculated by propagating messages from the candidate items $\mathcal{V}_{cand}$:
\begin{equation}\label{eq:s_p}
    s_p=g(u, p,\mathcal{V}_{cand}), 
\end{equation}
This adjacent attribute constraint brings two benefits. (1) In terms of recommendation, it significantly reduces the search space for selecting which attribute to ask.
Note that state-of-the-art conversational recommendation systems like EAR~\cite{lei20estimation} and CRM~\cite{Sun:2018:CRS:3209978.3210002} treat the whole attribute set $\mathcal{P}$ as the candidate space, increasing the difficulty to learn a good decision function. 
(2) In terms of conversation, constraining the adjacent attributes makes the dialogue more coherent. In linguistics, the closer the two entities are in any two adjacent utterances, the more coherent the conversation will be~\cite{gandhe2008evaluation}.

\subsubsection{Consultation} 
Once a reasoning step is completed, CPR moves to the \emph{consultation} step. The purpose of this step is to decide \emph{whether to ask an attribute} or \emph{to recommend items}, with the goal of achieving successful recommendations in fewest turns. We address it as a reinforcement learning~(RL) problem. Specifically, a policy function $\pi (\textbf{s})$ is expected to make the decision based on the global dialogue state $\textbf{s}$, which can include any information useful for successful recommendation, such as the dialogue history, the information of candidate items.
The output action space of the policy function contains two choices: $a_{ask}$ or $a_{rec}$, indicating whether to perform top-$k$ recommendations or to ask an attribute in this turn. If the RL decision is $a_{ask}$, we directly take highest-scored attribute from $\mathcal{P}_{cand}$, where the score is $s_p$ as defined in Eq. (\ref{eq:s_p}). Otherwise, we recommend top-$k$ items from $V_{cand}$ according to the score of $s_v$, which is defined in Eq. (\ref{eq:s_v}).

It is worth mentioning that our design of RL here reflects another major difference with existing conversational recommendation systems EAR~\cite{lei20estimation} and CRM~\cite{Sun:2018:CRS:3209978.3210002}.
Although they also learn policy networks with RL, their policy is to decide \textit{which attribute to ask}, rather than our choice of \textit{whether to ask attribute}. Which means, the size of their action space is $|\mathcal{P}| + 1$, where $|\mathcal{P}|$ denotes the number of attributes. 
This greatly increases the difficulty to learn the policy well, especially for a large $|\mathcal{P}|$, since RL is notoriously difficult to train when the action space is large~\cite{chen2019large}. In contrast, the action space of our RL is of size 2, being much easier to train.

\subsubsection{Transition} 
The \textit{transition} step will be triggered after the user confirms an asked attribute $p_{t}$. 
CPR first performs walking from the last confirmed attribute $p_{t-1}$ to $p_{t}$, forming an extended path $P=p_{0},p_{1},...p_{t-1},p_{t}$.
Then, we add $p_{t}$ to the preferred attribute set $\mathcal{P}_{u}$. Accordingly, the candidate attribute set is updated by $\mathcal{P}_{cand}=\mathcal{AA}_t \setminus (\mathcal{P}_{u} \cup \mathcal{P}_{rej})$, and the candidate item set $V_{cand}$ is updated by keeping the items that directly link to all attributes in the updated $\mathcal{P}_{u}$. Note that, if an attribute is rejected by the user, we just remove it from the candidate attribute set without vertex transition (line 9 of Algorithm \ref{algo:MCR}).
After the transition, our CPR starts the next conversation turn, repeating the same \emph{reasoning--consultation--transition} process
\footnote{Unlike EAR, CPR does not specifically answer the questions of \emph{how to adapt user feedbacks} like EARS by designing a reflection mechanism. It is because \cite{lei20estimation} reports it is still an open and challenging question with lots of details to be explored. Hence we leave it for future works.}
. 

\subsection{SCPR Model}

To materialize the CPR framework, we need to specify functions $f(v, u, \mathcal{P}_u)$, $g(u, p, \mathcal{V}_{cand})$ and $\pi(\textbf{s})$. We here provide a simple implementation SCPR, adapting some designs from EAR~\cite{lei20estimation} --- a latest conversational recommendation system.

\subsubsection{Reasoning - Item Scoring} In the reasoning stage, $f(v, u, \mathcal{P}_u)$ scores the item $v$ by propagating messages from the user-preferred attributes. 
We use the inner product between two vertex embeddings as the message, same as the FM variant used in EAR:
\begin{equation}
\begin{aligned}
    f(v,u,\mathcal{P}_u)= \mathbf{u}^T\mathbf{v} 
    + \sum_{p\in\mathcal{P}_u} \mathbf{v}^T \mathbf{p},
\end{aligned} 
\end{equation}
where $\textbf{u}$, $\textbf{v}$ and $\textbf{p}$ denote the embedding of the user $u$ and item $v$ and attribute $p$, respectively. 
The first term models the message propagated from the user to the item, and the second term models the messages propagated from user-preferred attributes to the item. These embeddings are randomly initialized and trained offline with the goal of scoring the interacted items higher than the non-interacted ones. 
The training objective is a multi-task pairwise loss, which follows EAR and we leave the details to Appendix~\ref{append-2}.

\subsubsection{Reasoning - Attribute Scoring}
Another function of the reasoning step is to decide which attribute is worth asking according to the current system state.
An expected strategy is to find the one that can better eliminate the uncertainty of items. 
As information entropy has proven to be an effective method of uncertainty estimation~\cite{wu2015probabilistic}, we implement the $g(u, p, \mathcal{V}_{cand})$ function as information entropy but adapt it to a weighted way:
\begin{equation}
\begin{aligned}
    g(u, p, \mathcal{V}_{cand}) &= -\text{prob}(p)\cdot \log_2(\text{prob}(p)), \\
    \text{prob}(p) &= \frac{\sum\limits_{v\in\mathcal{V}_{cand}\cap  \mathcal{V}_{p}} 
    \sigma(s_v) 
    }
    {\sum\limits_{v\in\mathcal{V}_{cand}} \sigma(s_v)},
\end{aligned}
\end{equation}
where $\sigma$ is the sigmoid function to normalize the item score $s_v$ to $(0,1)$, $\mathcal{V}_{cand}$ denotes the candidate items, and $\mathcal{V}_{p}$ denotes the items that include the attribute $p$. Different from the standard entropy which treats each item equally, our weighted entropy employed here assign higher weights to the important items (i.e., the items in $\mathcal{V}_p$ and scored higher) in attribute scoring. 
If there is no message propagated to an attribute, we define its entropy to be $0$. 
Note that, in this implementation, we do not consider user $u$ for calculating $g$ for simplicity. It does not not mean we don't value the importance of $u$ in deciding attribute. We leave the exploration of incorporating $u$ for future works.

\subsubsection{Consultation - RL Policy}
We use a two-layer feed forward neural network as our policy network. For the ease of convergence, we use the standard Deep Q-learning~\cite{mnih2015human} for optimization\footnote{According to our experiments, Deep Q-learning does not lead to better model, while making the model much easier to converge.
We also call the \textit{value network} (the terminology in Deep Q-learning) as policy network for ease of discussion.
}
 
The policy network takes the state vector $\textbf{s}$ as input and outputs the values ${Q(\textbf{s},a)}$ for the two actions, indicating the estimated reward for $a_{ask}$ or $a_{rec}$. A system will always choose the action with higher estimated reward.
The state vector $\textbf{s}$ is a concatenation of two vectors:
\begin{equation}\label{eq:state}
    \textbf{s} = 
    \textbf{s}_{his}
    \oplus
    \textbf{s}_{len},
\end{equation}
where $\textbf{s}_{his}$ encodes the conversation history, which is expected to guide the system to act smarter, e.g., if the asked attributes are accepted for multiple turns, it might be a suitable timing to recommend.
The $\textbf{s}_{len}$ encodes the size of candidate item set. As discussed by~\cite{lei20estimation}, it is easier to make successful recommendations when there are fewer candidate items. 

The reward follows~\cite{lei20estimation}, containing five kinds of rewards, namely, (1) $r_{rec\_suc}$, a strongly positive reward when the recommendation succeeds, (2) $r_{rec\_fail}$, a strongly negative reward when the recommendation fails, (3) $r_{ask\_suc}$, a slightly positive reward when the user accepts an asked attribute, (4) $r_{ask\_fail}$, a negative reward when the user rejects an asked attribute, and (5) $r_{quit}$, a strongly negative reward if the session reaches the maximum number of turns. The accumulated reward is the weighted sum of these five. The detailed value for each reward can be found in {\color{black}Sec~\ref{sec: exp-setup}}. \vspace{+5pt}

\noindent While some components of our SCPR are adapted from EAR, it is worth highlighting two significant differences between them.
First, SCPR leverages on the \emph{adjacent attribute} constraint on the graph, largely reducing the search space of attributes. 
Second, SCPR scores attributes through message propagation on the graph, instead of by the policy network as what has been done in EAR. 
This enables our policy network to have a \emph{much smaller decision space} --- only two actions, alleviating the pressure for policy making.

\section{Experiments}
In this section, we are going to evaluate our proposed CPR framework by empirically examining the SCPR implementation on two real-world datasets.
We use the following research questions (RQs) to guide our experiment\footnote{Code and datasets can be found at: https://cpr-conv-rec.github.io/}.
\begin{itemize}[leftmargin=1mm]
	\item \textbf{RQ1.} How does our CPR framework compared with existing conversational recommendation methods?
    \item \textbf{RQ2.} Are the \emph{adjacent attribute constraint} and \emph{smaller decision space} in SCPR really effective?
	\item \textbf{RQ3.} Can our method make the reasoning path explainable and easy-to-interpret?
\end{itemize}
\subsection{Dataset Description}
For better comparison, we follow EAR~\cite{lei20estimation} to conduct experiments on LastFM\footnote{ https://grouplens.org/datasets/hetrec-2011/} for music artist recommendation and Yelp\footnote{https://www.yelp.com/dataset/} for business recommendation. 
LastFM contains 1,801 users and 27,675 items and 76,693 interactions. Yelp contains 27,675 users, 70,311 items and 1,368,606 interactions. 

In the original paper of EAR~\cite{lei20estimation}, LastFM is designed to evaluate binary question scenario, where the user give preference towards an attribute using yes or no. For the ease of modeling, \citet{lei20estimation} manually merged relevant attributes into 33 coarse-grained attributes.
Whereas, the Yelp dataset is designed for enumerated questions, where the user can select multiple attributes under one category. They manually built a 2-layer taxonomy and there are 29 first-layer categories with 590 second-layer attributes.

While we follow the setting of \cite{lei20estimation}, we believe they are not necessarily the best practice as they requires heavy manual efforts with expert knowledge, which is expensive for real usage. Therefore we also consider the setting of using the original attributes (pruning off frequency < 10 attributes), denoting them as LastFM* (containing 8438 attributes) and Yelp* (containing 590 attributes) separately. The statistics can be found in Appendix \ref{append-1}.

\subsection{Experimental Setup}
\label{sec: exp-setup}
\subsubsection{Training Details}
We split each dataset for training, validation and testing in a ratio of 7:1.5:1.5. And set top k item as 10, and maximum turn T as 15.
Following~\cite{lei20estimation,Sun:2018:CRS:3209978.3210002} The training process is made up of two parts: (1) An offline training for scoring function of item in \emph{reasoning} step. We use the historical clicking record in the training set to optimize our factorization machine offline (Eq.~(3)) by strictly follow~\cite{lei20estimation}. The goal is to assign higher score to the clicked item for each users. We articulate the details in Appendix \ref{append-2} and we also refer the readers to the original paper~\cite{lei20estimation} for more Information.
All hyperparameters for offline training remains the same as~\cite{lei20estimation}.
(2) An online training for reinforcement learning used in consultation step. We use a user simulator (\emph{c.f.} Sec~\ref{sec:simu}) to interact with the user to train the policy network using the validation set. 
The detailed rewards to train the policy network are: $r_{rec\_suc}$=1, $r_{rec\_fail}$=-0.1, $r_{ask\_suc}$=0.01, $r_{ask\_fail}$=-0.1, $r_{quit}$=-0.3. The parameters of the DQN are empirically set as following: the experience replay memory size is 50,000, the sample batch size is 128, discount factor $\gamma$ is set to be 0.999. We optimize policy network with RMSprop optimizer and update the target network every 20 epsiodes. It is also noticeable that, since our policy network have much smaller actions space and we adopt Deep Q-learning, the network is easier to converge. We do not need to have pre-training as adopted in EAR~\cite{lei20estimation} and CRM~\cite{Sun:2018:CRS:3209978.3210002}.
All those hyperparameters related online training are tuned according to the validation set.

\subsubsection{User Simulator For MCR}
\label{sec:simu}
As a CRS is an interactive system, it needs to be trained and evaluated by interacting with users. However, it is infeasible to do so in a research lab. Employing a user simulator is a common practice~\cite{chandramohan2011user}. We follow the 
user simulators in~\cite{Sun:2018:CRS:3209978.3210002, lei20estimation} which simulate one conversation session for one user-item $(u, v)$ interaction record in validation set (for training) and testing set (for testing). In a giving session, the user $u$'s preference is anchored by item $v$: (1) when the system proposes a list of items, he will only accept it if the list contains item $v$; (2) when a system asks for an attribute, he will only confirm he likes it if this attribute is included by item $v$.
There is no denying that such simulation has many limitation, but it is the most practical and realistic at current stage~\cite{Sun:2018:CRS:3209978.3210002, lei20estimation}. 
One major attack for such simulation is that the user may "falsely" reject an item which is actually liked by him but it has not been observed hence not being clicked by him. However, it is hard to address it as there is few suitable exposure data.
One may also suggest to treat all user-item interaction in testing set as positive instances for one session, we also forgo using it because the aim for CRS is to capture user's current specific preference which may shift from his general interest. As our main focus is the strategy of graph reasoning, we use template for conversations.

\subsubsection{Baselines}
Although there are more CRS models, they follow different settings, hence being not comparable to us. We use the following baselines to compare:
\begin{itemize}[leftmargin=1mm]
    \item \textbf{Max Entropy}. This method follows a rule-based protocol to ask and recommend. When asking question, it always chooses an attribute with the maximum entropy within the current candidate item set. The system follows certain probabilities to recommend. Details can be found at~\cite{lei20estimation}.
    \item \textbf{Abs Greedy}~\cite{christakopoulou2016towards}. This method serves as a baseline where the model only recommends items and updated itself, until it finally makes successful recommendation.
    \citet{christakopoulou2016towards} report that it outperforms online recommendation methods like Thompson Sampling~\cite{chapelle2011empirical}.
    \item \textbf{CRM}~\cite{Sun:2018:CRS:3209978.3210002}. This is a CRS model which records user's preference into a belief tracker, and uses reinforcement learning (RL) to find the policy to interact with the user. The RL leverages on a policy network whose state vector is the result of belief tracker. We follow~\cite{lei20estimation} to adapt it to the MCR scenario.
    \item \textbf{EAR}~\cite{lei20estimation}. This is the state-of-the-art method on MCR setting and proposed a three stage solution called Estimation--Action--Reflection which emphasizes on the interaction between conversation component and recommendation component. This inspires our SCPR implementation hence being the most comparable model.

\end{itemize}

\subsubsection{Evaluation Metrics}
The evaluation follows~\cite{lei20estimation}. We use success rate (SR@t)~\cite{Sun:2018:CRS:3209978.3210002} to measure the cumulative ratio of successful recommendation by turn $t$.
We also use average turns (AT) to record the average number of turns for all session (if a session still fails in the last turn $T$, we count the turn for that session as $T$). Therefore, the higher SR@t indicates a higher performance at a specific turn t, while the lower AT means an overall higher efficiency.

\subsection{Performance Comparison of SCPR with Exsiting Models (RQ1)}
\label{sec: main-result}
\begin{table}[htbp]

  \centering
  \caption{Success Rate @ 15 and Average Turn. \textbf{Bold number} represents the improvement of SCPR over existing models is statistically significant ($p<0.01$) (RQ1)}
    \begin{tabular}{c|c|c|c|c|}
    \hline
          & \multicolumn{2}{c|}{\textbf{LastFM}} & \multicolumn{2}{c|}{\textbf{Yelp}} \\
    \hline
          & \textbf{SR@15} & \textbf{AT} & \textbf{SR@15} & \textbf{AT} \\
    \hline
    Abs Greedy & 0.222  & 13.48  & 0.264  & 12.57  \\
    \hline
    Max Entropy & 0.283  & 13.91   & 0.921  &  6.59  \\
    \hline
    CRM   & 0.325 & 13.75  & 0.923 & 6.25  \\
    \hline
    EAR   & 0.429 & 12.88  & 0.967 & 5.74 \\
    \hline
    SCPR   & \textbf{0.465} & \textbf{12.86} & \textbf{0.973} & \textbf{5.67}\\
    \hline
    \end{tabular}%
  \label{tab:RQ1}%
  \vspace{-15pt}
\end{table}%

\begin{table}[htbp]
  \centering \vspace{-5pt}
  \caption{Performance comparison on original attributes. Bold number represents the improvement of SCPR over existing models is statistically significant ($p<0.01$) (RQ1)}
    \begin{tabular}{c|c|c|c|c|}
    \hline
          & \multicolumn{2}{c|}{\textbf{LastFM*}} & \multicolumn{2}{c|}{\textbf{Yelp*}} \\
    \hline
          & \textbf{SR@15} & \textbf{AT} & \textbf{SR@15} & \textbf{AT} \\
    \hline
    Abs Greedy & 0.635  & 8.66  & 0.189   & 13.43   \\
    \hline
    Max Entropy & 0.669   & 9.33  &  0.398  & 13.42   \\
    \hline
    CRM   & 0.580  & 10.79  & 0.177 & 13.69  \\
    \hline
    EAR   &   0.595  & 10.51  & 0.182  & 13.63 \\
    \hline
    SCPR   &  \textbf{0.709}  & \textbf{8.43} & \textbf{0.489} & \textbf{12.62}\\
    \hline
    \end{tabular}%
  \label{tab:RQ2}%
  \vspace{-10pt}
\end{table}%

\begin{figure}[htbp]
\begin{minipage}[t]{0.47\linewidth}
    \includegraphics[width=\linewidth]{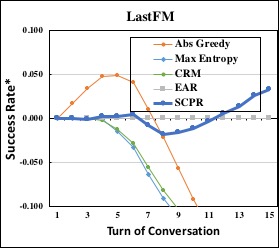}
    \label{f1}
\end{minipage}%
    \hfill%
\begin{minipage}[t]{0.47\linewidth}
    \includegraphics[width=\linewidth]{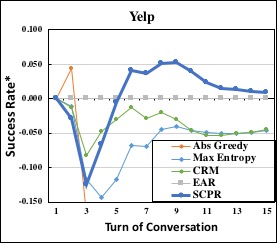} 
    \label{f2}
\end{minipage}
\vspace{-15pt}
\caption{Success Rate* of compared methods at different turns on LastFM and Yelp (RQ1).
} 
\label{fig:overall}
\end{figure}

\begin{figure}[htbp]
\begin{minipage}[t]{0.47\linewidth}
    \includegraphics[width=\linewidth]{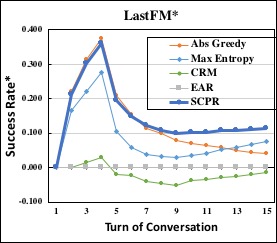}
    \label{f1}
\end{minipage}%
    \hfill%
\begin{minipage}[t]{0.47\linewidth}
    \includegraphics[width=\linewidth]{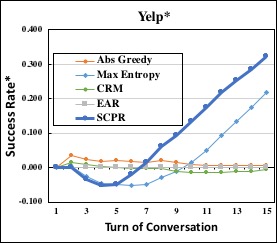} 
    \label{f2}
\end{minipage}
\vspace{-15pt}
\caption{Success Rate* of compared methods at different conversation turns on LastFM* and Yelp*(RQ1).} 
\label{fig:overall-star}
\end{figure}

\begin{figure*}[t]
    \centering
	\includegraphics[height=5cm]{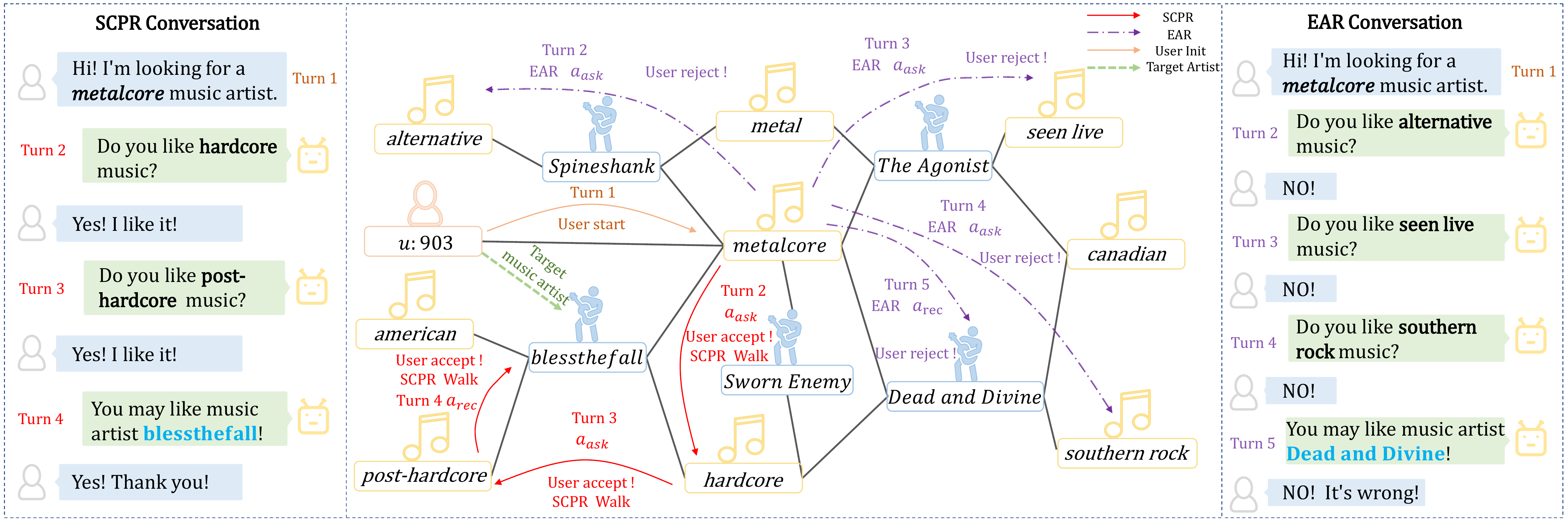}
    \caption{Sample conversations generated by SCPR (left) and EAR (right) and their illustrations on the graph (middle). 
    } 
    \label{fig:case-study}
    \vspace{-10pt}
\end{figure*}
\vspace{5pt}
Table \ref{tab:RQ1} and \ref{tab:RQ2} present the statistics of model's performances. We can see that our SCPR model achieves significantly higher SR and less AT than state-of-the-art baselines, demonstrating our SCPR method's supurior performances in usage.

We also intuitively present the performance comparison in Figure \ref{fig:overall} and \ref{fig:overall-star}.
They show Success Rate * (SR*) in each turn. SR* denotes the relative SR compared with the most competitive baseline EAR (meaning the difference of SR between each method and EAR), and EAR serves as the gray line of y = 0 in the figures.
We have following discoveries:
\begin{itemize}[leftmargin=1mm]
    \item It is important to see our SCPR outperforms all baselines on various settings. Interestingly, we can find that our SCPR shows larger advantage in LastFM* and Yelp* datasets, showing its validity in practical usage with large attribute space. 
    This also validates our key design in CPR. Firstly, the graph constraint helps our model to eliminate many irrelevant attributes to ask, this becomes especially helpful when there are a large number of attributes. In contrast, we discover that EAR has more difficulties in asking accurate attributes in LastFM* and Yelp*.
    Secondly, our framework utilizes a more dedicated RL model which only decides to recommend or to ask, hence have better chances to learn more effective policy. 
    At the same time, EAR may outperform SCPR on first few rounds, but it falls behind in future rounds. It is due to EAR has much larger action space, making it challenging for them to learn effective strategy to recommend while asking.

    \item Interestingly, Abs Greedy can achieve the best results on the first few turns but plunges in further turns. The reason is that Abs Greedy is the only method that solely attempts to recommend items to user.
    While Abs Greedy is continuously pushing recommendation, other methods are probably consulting user's explicit feedback on attributes which in turns helps reduce candidate item space and help the model achieve long term reward. This also validates our core design in SCPR -- utilizing user's explicit feedback on attributes.
    
    \item {\color{black}The two previously proposed RL-based methods, EAR and CRM can both outperform max entropy in Yelp and LastFM, but achieve lower performance than max entropy in Yelp* and LastFM*.
    The reason is that Yelp* and LastFM* have larger attribute space (590 and 8438 than 29 and 33).
    According to their model design, their RL model is responsible for both \textit{which attribute to ask} and \textit{whether to recommend}. Therefore, they have action spaces of 590+1 dimensions and 8438+1 dimensions for Yelp* and LastFM* respectively. Such larger action space may bring challenges to action making.}

\end{itemize}

\vspace{-8pt}
\subsection{Evaluating Key Design in SCPR (RQ2)}
The key design of our SCPR method is that we leverage on the \textit{adjacent attribute constraint} of the graph and a more dedicated RL model with smaller action space. To test the effectiveness of such key features, we conduct additional experiments by designing a variant of our SCPR model, named SCPR-v. Specifically, we replace our policy network with the policy network in EAR. It has the same state vector as EAR and the action space of the policy network increases from 2 to $|\mathcal{P}|+1$, meaning that the policy function is also responsible for \textit{deciding which attribute to ask}.
Note that we keep other components unchanged, including our graph constraint of adjacent attributes. 
Such constraint exerts influence on our policy function in a straightforward way: we add a condition of "being one adjacent attribute" to the selection of action with maximum value.
Therefore such SCPR-v model can be seen as an intermediate layer between EAR and SCPR, (1) it can be seen as a variant of SCPR where its RL model is not so dedicated, (2) it can also be seen as a variant of EAR where the attribute asking can be helped (if any) by our graph constraint. We follow the same implementation paradigm for SCPR-v. Due to the space limitation, we only briefly report the Success Rate* comparison among SCPR-v, EAR and SCPR on LastFM* and Yelp* datasets which are more representative.

\begin{figure}[htbp]
\begin{minipage}[t]{0.47\linewidth}
    \includegraphics[width=\linewidth]{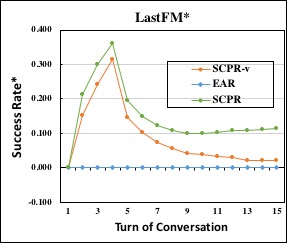}
    \label{f1}
\end{minipage}%
    \hfill%
\hspace{+0.1pt}
\begin{minipage}[t]{0.47\linewidth}
    \includegraphics[width=\linewidth]{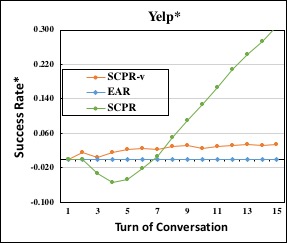} 
    \label{f2}
\end{minipage}
\vspace{-15pt}
\caption{Success Rate* of compared methods at different conversation turns on LastFM* and Yelp*(RQ2).} 
\label{fig:ablation}
\end{figure}

We have these discoveries: 
(1) SCPR-v has generally worse performance than SCPR, since it is very challenging for decision making in a very large action space. It validates our design in a more dedicated RL model with small action space. 
Interestingly we can see that SCPR-v has similar performance at first few turns compared with SCPR, but falls behind in future turns.
According to instance-level studies, we observe the RL component of SCPR-v adopts simple strategies. It asks a few attributes and recommend items at earlier turns than SCPR.
Then with fewer attributes known, it has a larger candidate item space, making it harder to achieve higher SR in longer turns.
It suggests that SCPR-v with very large action space has difficulty learning effective strategy for long term reward.
(2) We can see that SCPR-v has better performance than EAR. This advantage is due to our graph constraint on attributes that eliminates many irrelevant attributes, making it easier for attribute choice.

\vspace{-10pt}
\subsection{Case Study on Explainability (RQ3)}

Aside from the superior performance on success rate and average turns, our CPR is also more explainable. It conducts conversational recommendation by walking (reasoning) on graph, resulting in a path of attributes. 
The path brings crystally clear reasoning logic which is naturally explainable. 

Let's go through an example of real interaction from LastFM* in Figure~\ref{fig:case-study}. We display the conversation histories of SCPR and EAR on two sides of the figure, and illustrate the whole processing in the middle graph.
The session is initiated by the user (id: 903) who specifies an attribute he likes as $``metalcore"$. We can see our SCPR travels a short path of attributes ($``metalcore"$ to $``hardcore"$ then to $``post$-$hardcore"$) that quickly reaches user's preferred item (artist $``blessthefall"$) and successfully make recommendation. The whole conversation is coherent and the red path is the explanation of the recommendation reason. On the contrary, EAR's behavior looks strange. It first asks $``alternative"$, then asks $``seenlive"$ followed by $``southern \ rock"$. Those attributes are not very closely related, being more like a random pop-ups of attributes. From model developing perspective, such loss of relevance makes it hard to explain why the EAR executes such jumps. From application perspective, such loss of relevance leads to less coherent conversations.

\section{Conclusion and Future Work}
We are the first to introduce graph to address the multi-round conversational recommendation problem, and propose the \emph{Conversational Path Reasoning} (CPR) framework. 
CPR synchronizes conversation with the graph-based path reasoning, making the utilization of attribute more explicitly hence greatly improving explainability for conversational recommendation. 
Specifically, it tackles \emph{what item to recommend} and \emph{what attribute to ask} problems through message propagation on the graph, leveraging on the complex interaction between attributes and items in the graph to better rank items and attributes. Using the graph structure, a CRS only transits to the adjacent attribute, reducing the attribute candidate space and also improving the coherence of the conversation.
Also, since the items and attributes have been ranked during the message propagation, the policy network only needs to decide \emph{when to ask and when to recommend}, reducing the action space to be $2$. It relieves the modeling load of the policy network, enabling it to be more robust especially when the candidate space is large. 

There are many interesting problems to be explored for CPR. First, CPR framework itself can be further improved. For example, CPR does not consider how to adapt the model when the user rejects a recommended item. How to effectively consider such rejected items would be an interesting and challenging task.
Second, more sophisticated implementation can be considered. For example, it is possible to build more expressive models for attribute scoring other than the weighted max-entropy as adopted in this paper. Currently, the embeddings of items and attributes do not get updated during the interactive training. It would be better to build a more holistic model to incorporate the user feedback to update all parameters in the model, inclusive of user, item and attribute embeddings.

\textbf{Acknowledgement}:
This research is supported by the National Research Foundation, Singapore under its International Research Centres in Singapore Funding Initiative as well as National Natural Science Foundation of China (61972372, U19A2079). All content represents the opinion of the authors, which is not necessarily shared or endorsed by their respective employers and/or sponsors.
We thank the anonymous reviewers for their valuable comments.

\bibliographystyle{ACM-Reference-Format}

\bibliography{reference.bib}

\newpage

\appendix
\newpage
\newpage
\section{Dataset Statistics}
\label{append-1}
\begin{table}[htbp]
  \centering
  \caption{Dataset Statictics of LastFM and Yelp. Here we list the relation types in different datasets to let readers to get better understanding of the dataset.} 
    \begin{tabular}{|c|c|c|c|}
    \toprule
    \multicolumn{2}{|c|}{\textbf{Dateset}} & \textbf{LastFM} & \textbf{Yelp} \\
    \midrule
    \hline
    \multicolumn{1}{|c|}{\multirow{4}[2]{*}{\makecell{User-Item \\ Interaction}}} & \#Users & 1,801  & 27,675  \\
          & \#Items & 7,432  & 70,311  \\
          & \#Interactions & 76,693  & 1,368,606  \\
          & \#attributes & \textbf{33}    & \textbf{29}  \\
    \midrule
    \hline
    \multicolumn{1}{|c|}{\multirow{3}[2]{*}{Graph}} & \#Entities & 9,266  & 98,605  \\
          & \#Relations & 4     & 3  \\
          & \#Triplets & 138,217  & 2,884,567  \\
    \midrule
    \hline
    \textbf{Relations} & \textbf{Description} & \multicolumn{2}{c|}{\textbf{Number of Relations}} \\
    \midrule
    Interact & user $\Longleftrightarrow{}$ item & 76,696  & 1,368,606  \\
    Friend & user $\Longleftrightarrow{}$ user & 23,958  & 688,209  \\
    Like  & user $\Longleftrightarrow{}$ attribute & 7,276  & * \\
    Belong\_to & item $\Longleftrightarrow{}$ attribute & 30,290  & 350,175  \\
    \bottomrule
    \end{tabular}%
    \label{tab:basic}
\end{table}%

\begin{table}[htbp]
  \centering
  \caption{Dataset Statistics for LastFM* and Yelp*, we use original attributes to avoid complex feature engineering.}
    \begin{tabular}{|c|c|c|c|}
    \toprule
    \multicolumn{2}{|c|}{\textbf{Dateset}} & \textbf{LastFM*} & \textbf{Yelp*} \\
    \midrule
    \hline
    \multicolumn{1}{|c|}{\multirow{4}[2]{*}{\makecell{User-Item \\ Interaction}}} & \#Users & 1,801  & 27,675  \\
          & \#Items & 7,432  & 70,311  \\
          & \#Interactions & 76,693  & 1,368,606  \\
          & \#attributes & \textbf{8,438}  & \textbf{590}  \\
    \midrule
    \hline
    \multicolumn{1}{|c|}{\multirow{3}[2]{*}{Graph}} & \#Entities & 17,671  & 98,576  \\
          & \#Relations & 4     & 3  \\
          & \#Triplets & 228,217  & 2,533,827  \\
    \midrule
    \textbf{Relations} & \textbf{Description} & \multicolumn{2}{c|}{\textbf{Number of Relations}} \\
    \midrule
    Interact & user $\Longleftrightarrow{}$ item & 76,696  & 1,368,606  \\
    Friend & user $\Longleftrightarrow{}$ user & 23,958  & 688,209  \\
    Like  & user $\Longleftrightarrow{}$ attribute & 33,120  & * \\
    Belong\_to & item $\Longleftrightarrow{}$ attribute & 94,446  & 477,012  \\
    \bottomrule
    \end{tabular}%
  \label{tab:original-attribute}%
\end{table}%

\newpage
\section{Details of Offline training in the \textit{Reasoning} Step}
\label{append-2}
In our \textbf{CPR} framework design, there is a trainable component in \textit{reasoning} step for item scoring. For simplicity, we instantiate it as the FM model in EAR\cite{lei20estimation}. For the reproducibility of this paper, we articulate the whole process of such instantiation in this section.

\subsection{Training Objective}
EAR\cite{lei20estimation} embeds users, items and attributes as vectors into one Factorization Machine (FM)\cite{rendle2010factorizing} model. The training objective for such FM model is simultaneously achieving item prediction and attribute prediction for Multi-round Conversational Recommendation(MCR) scenario, using a multi-task pairwise loss.

\subsubsection{Item Prediction}
We borrow a variant of Factorization Model (FM) model as introduced in \cite{lei20estimation} to capture the interaction between users, items and attributes. As discussed in Section \ref{sec:CPR_framework}, the scoring function is defined as:
\begin{equation}
\begin{aligned}
    f(u,v,\mathcal{P}_u)= \mathbf{u}^T\mathbf{v}
    + \sum_{p_i\in\mathcal{P}_u} \mathbf{v}^T \mathbf{p_i},
\end{aligned}
\end{equation}
where the first and second term represent message propagated from user to item and item to user respectively.

We follow \cite{lei20estimation} to use a pairwise loss to optimize,
one key innovation is that they use two types of negative samples $\mathcal{D}_1$ and $\mathcal{D}_2$ tailored for MCR:
\begin{equation}
\begin{aligned}
    L_{item} &= \sum_{(u,v,v')\in \mathcal{D}_1} -\mbox{ln} \sigma(f(u,v,\mathcal{P}_u) - f(u,v',\mathcal{P}_u)) \\
    &+ \sum_{(u,v,v')\in \mathcal{D}_2} -\mbox{ln} \sigma(f(u,v,\mathcal{P}_u) - f(u,v',\mathcal{P}_u)) + \lambda_{\Theta} \left\|\Theta\right\|^2,
\end{aligned}
\end{equation}
where
\begin{equation}
    \mathcal{D}_1 := \{(u, v, v') \mid v' \in \mathcal{V}_{u}^- \},
    \quad
    \mathcal{V}_{u}^{-} := \mathcal{V} \backslash \mathcal{V}_{u}^{+}
\end{equation}
\begin{equation}
\begin{aligned}
    \mathcal{D}_2 := \{(u, v, v') \mid v' \in \widehat{\mathcal{V}}_{u}^- \},
    \quad
    \widehat{\mathcal{V}}_{u}^- := \mathcal{V}_{cand} \backslash \mathcal{V}_{u}^{+}
\end{aligned}
\end{equation}

The intuition is that the model first needs to learn user's general preference ($\mathcal{D}_1$). Additionally it also should learn user's preference when some attributes have been confirmed, resulting a dynamically updating candidate item set $\mathcal{V}_{cand}$, which is a main characteristic for MCR ($\mathcal{D}_2$).
Specifically, $\mathcal{V}_{u}^{-}$ is the ordinary negative samples which are non-interacted items, and $\widehat{\mathcal{V}}_{u}^-$ is candidate item set $\mathcal{V}_{cand}$ that excludes the interacted items $\mathcal{V}_{u}^{+}$. The obtaining of $\mathcal{V}_{cand}$ is a dynamic process, which will be discussed in Section \ref{sec:data collection}.

\subsubsection{Attribute Prediction}
The EAR\cite{lei20estimation} also leverages on the FM model to make attribute prediction. Intuitively, the next attribute $p$ to ask should be dependent on the confirmed attribute set $\mathcal{P}_u$, formally:

\begin{equation}\label{eq:sim}
    \hat{g}(p|u,\mathcal{P}_u) = \textbf{u}^T \textbf{p} + \sum_{p_i\in \mathcal{P}_u} \textbf{p}^T \textbf{p}_i,
\end{equation}
where the first term captures user's general preference towards the given attribute $p$, and the second term models the interaction between $p$ and each attribute in the confirmed attribute set $\mathcal{P}_u$.

They similarly leverages on the pairwise loss for attribute prediction:
\begin{equation}
\begin{aligned}
    L_{attr} = \sum_{(u,p,p')\in \mathcal{D}_3} -\mbox{ln} \sigma(\hat{g}(p|u, \mathcal{P}_u) - \hat{g}(p'|u, \mathcal{P}_u)) + \lambda_{\Theta} \left\|\Theta\right\|^2,
\end{aligned}
\end{equation}
where the pairwise training data $\mathcal{D}_3$ is defined as:
\begin{equation}
    \mathcal{D}_3 = \{(u, p, p')| p\in \mathcal{P}_v, p'\in \mathcal{P} \backslash \mathcal{P}_v \},
\end{equation}
The $\mathcal{P}_v$ here denotes attributes of item $v$, hence $p$ and $p'$ represent attributes that belongs and \textit{not} belongs to current item respectively, forming the pairwise sample.

\subsubsection{Multi-task learning}
Since \cite{lei20estimation} discovered that item prediction and attribute prediction can mutually promote, we also follow their practice to use such multi-task pairwise loss to achieve these two goals:
\begin{equation}
    L = L_{item} + L_{attr}.
\end{equation}

\subsection{Data Collection}
\label{sec:data collection}
As we have elaborated the training objective of the FM model used in the \textit{reasoning} step, now we are going to describe how we obtain the data used to train such model, which are in fact $\mathcal{D}_1$, $\mathcal{D}_2$ and $\mathcal{D}_3$.

As a common practice introduced in Section \ref{sec:simu}, we also leverage on user simulator to obtain such data. As introduced before, we use observed user-item interactions to ground such simulation.
We accumulate $\mathcal{D}_1$, $\mathcal{D}_2$ and $\mathcal{D}_3$ through many MCR sessions and append new instances at each steps of the interactions.
Specifically, given a user $u$ and an item $v$ which has an attribute set $\mathcal{P}_v$, without the loss of generality, we assume $\mathcal{P}_v = \{p_0, p_1, p_2, p_3, p_4\}$.
As for the accumulation of $\mathcal{D}_1$, it is actually independent from the interaction steps because it is intrinsically static. Therefore we directly \vfill\eject \noindent sample one item from the non-interacted items of user $u$.
Now let's assume we are at the stage where user has confirmed a few attributes, yielding the confirmed attribute set $\mathcal{P}_u = \{p_0, p_1, p_2\}$. Now $\mathcal{V}_{cand}$ is the set of items satisfying all attributes in $\mathcal{P}_u$, one negative instance will be sampled from the non-interacted items in $\mathcal{V}_{cand}$ to form $\mathcal{D}_2$.
Note that the positive instances in pairwise samples are always $v$ for both $\mathcal{D}_1$ and $\mathcal{D}_2$.
Finally, as for the attribute side, the positive instances for attributes are $\{p_3, p_4\}$, each of them will be paired with a negative instance sampled from
$\mathcal{P} \backslash \mathcal{P}_v$ and be added into $\mathcal{D}_3$.

In order to have a high coverage of the dataset, we use all user-item interactions in training set to ground such simulation. What's more, we simulate multiple times for each user-item interaction, with all possibility of the first attribute user informed $p_0$ being tried.

\subsection{Training Details}
After the training data has been collected, we strictly follow the training instruction in \cite{lei20estimation}.
To briefly report, we set the embedding size of FM model as 64. We used SGD optimizer with L2 regularization of 0.001. The learning rate for item prediction task and attribute prediction task are set us 0.01 and 0.001 repectively.

\end{document}